# Anderson localization of a non-interacting Bose-Einstein condensate

G. Roati, C. D'Errico, L. Fallani, M. Fattori[1], C. Fort, M. Zaccanti, G. Modugno, M. Modugno[2] & M. Inguscio

*LENS and Physics Department, Università di Firenze, and INFM-CNR, Via Nello Carrara 1, 50019 Sesto Fiorentino, Italy*

[1]*Museo Storico della Fisica e Centro Studi e Ricerche "E. Fermi", Roma, Italy*
[2]*Dipartimento di Matematica Applicata, Università di Firenze, Italy – BEC-INFM Center, Univ. Trento, Italy*

**One of the most intriguing phenomena in physics is the localization of waves in disordered media[1]. This phenomenon was originally predicted by Anderson, fifty years ago, in the context of transport of electrons in crystals[2]. Anderson localization is actually a much more general phenomenon[3], and it has been observed in a large variety of systems, including light waves[4,5]. However, it has never been observed directly for matter waves. Ultracold atoms open a new scenario for the study of disorder-induced localization, due to high degree of control of most of the system parameters, including interaction[6]. Here we employ for the first time a non-interacting Bose-Einstein condensate to study Anderson localization. The experiment is performed with a one-dimensional quasi-periodic lattice, a system which features a crossover between extended and exponentially localized states as in the case of purely random disorder in higher dimensions. Localization is clearly demonstrated by investigating transport properties, spatial and momentum distributions. We characterize the crossover, finding that the critical disorder strength scales with the tunnelling energy of the atoms in the lattice. Since the interaction in the condensate can be controlled at will, this system might be employed to solve open questions on the interplay of disorder and interaction[7] and to explore exotic quantum phases[8,9].**

The transition between extended and localized states originally studied by Anderson for non-interacting electrons has actually never been observed in crystals due to the high electron-electron and electron-phonon interactions[2]. Researchers have therefore turned their attention to other systems, where interactions or nonlinearities are almost absent. Evidence of the Anderson localization for light waves in disordered media has been provided by an observed modification of the classical diffusive regime, featuring a conductor-insulator transition[4,5]. However, clear understanding of the interplay between disorder and nonlinearity is considered a crucial task in contemporary physics. First effects of weak nonlinearities have been recently shown in experiments with light waves in photonic lattices[10,11]. The combination of ultracold atoms and optical potentials is offering a novel platform to study disorder-related phenomena, where most of the relevant physical parameters, including interaction, can be controlled[6,8]. The introduction of laser speckles[12] and quasi-periodic optical lattices[9] have made possible the investigation of the physics of disorder. The investigations reported so far have explored either quantum phases induced by interaction[9] or regimes of weak interaction where however the observation of Anderson localization was precluded either by the size of disorder or by delocalizing effects of nonlinearity[12-16].

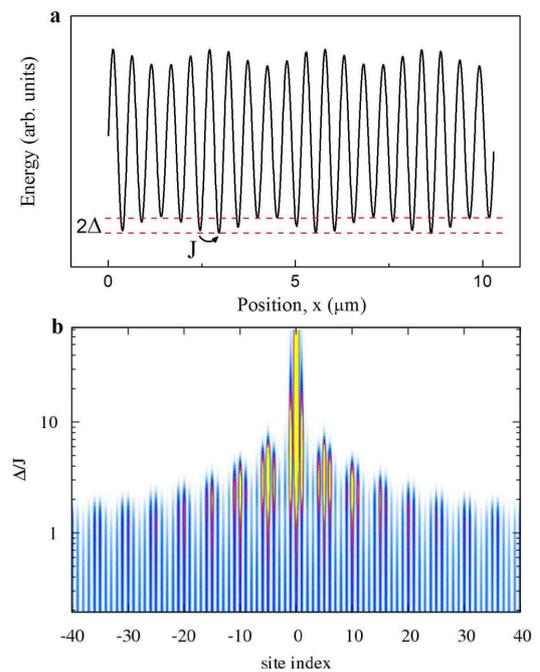

**Figure 1 | The quasi-periodic optical lattice. a,** Sketch of the quasi-periodic potential realized in the experiment. The hopping energy J describes the tunnelling between different sites of the primary lattice and $2\Delta$ is the maximum shift of the on-site energy induced by the secondary lattice. The lattice constant is 516 nm. **b,** Typical density plot of an eigenstate of the bichromatic potential, as a function of $\Delta/J$ (vertical axis). For small values of $\Delta/J$ the state is delocalized over many lattice sites. For $\Delta/J \geq 7$ the state becomes exponentially localized on lengths smaller than the lattice constant.

In this work, we employ for the first time a Bose-Einstein condensate where the interaction can be tuned independently from the other parameters[17], to study localization purely due to disorder. We study localization in a one-dimensional (1D) lattice perturbed by a second, weak incommensurate lattice, which constitutes the first experimental realization of the non-interacting Harper[18] or Aubry-André model[19]. This quasi-periodic system displays a transition from extended to localized states analogous to the Anderson transition, already in 1D[20,21] whereas in case of pure random disorder, dimensions higher than two would be needed[22]. We clearly observe this transition by studying transport, and both spatial and momentum distributions, and we verify the scaling behaviour of the critical disorder strength.

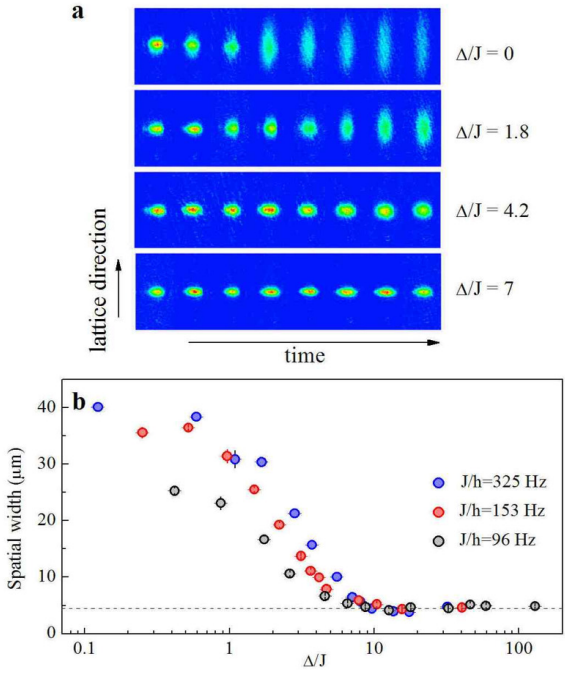

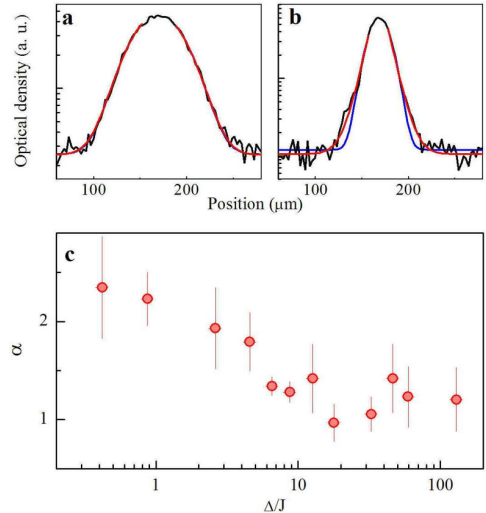

[16]. The Gaussian shape of the laser beams forming the primary lattice provides also radial confinement of the condensate in absence of the harmonic trap.

In a first experiment we have investigated transport, by abruptly switching off the main harmonic confinement and letting the atoms expand along the 1D bichromatic lattice. We detect the spatial distribution of the atoms after increasing evolution times by absorption imaging, see Fig.2a. In a regular lattice ($\Delta/J=0$) the eigenstates of the potential are extended Bloch states, and the system expands ballistically. In the limit of large disorder ($\Delta/J\gtrsim 7$) we observe no diffusion, since in this regime the condensate can be described as the superposition of several localized eigenstates, whose individual extension is smaller than the initial size of the condensate. In the crossover between these two regimes we observe a ballistic expansion with reduced speed. This crossover is summarized in Fig.2b, where we report the width of the atomic distribution at a fixed evolution time of 750 ms vs the rescaled disorder strength $\Delta/J$, for three different values of J. In all three cases, the system enters the localized regime at the same disorder strength, providing a compelling evidence of the scaling behaviour intrinsic in the model (1).

**Figure 2 | Probing the localization with transport. a**, In-situ absorption images of the BEC diffusing along the quasi-periodic lattice for different values of $\Delta$ and J/h=153 Hz. For $\Delta/J\gtrsim 7$ the size of the BEC remains stacked to its original value, reflecting the onset of localization. **b**, Rms size of the condensate for three different values of J, at a fixed diffusion time of $\tau$ =750 ms, vs the rescaled disorder strength $\Delta/J$. The dashed line indicates the initial size of the condensate. The onset of localization appears in all three cases in the same range of values of $\Delta/J$.

Our system is described by the Aubry-André hamiltonian:
$$H = J\sum_{m}(|w_m\rangle\langle w_{m+1}|+|w_{m+1}\rangle\langle w_m|) + \Delta\sum_{m}\cos(2\pi\beta m+\phi)|w_m\rangle\langle w_m| \quad (1)$$
where $|w_m\rangle$ is the Wannier state localized at the lattice site $m$, J is the site-to-site tunnelling energy, $\Delta$ is the strength of disorder and $\beta=k_2/k_1$ is the ratio between the two lattice wavevectors. In the experiment, the two relevant energies J and $\Delta$, can be controlled independently, by changing the height of the primary and secondary lattice potential, respectively. For a maximally incommensurate ratio $\beta=(\sqrt{5}-1)/2$, the model exhibits a sharp transition from extended to localized states at $\Delta/J =2$ [18,19,21]. For the actual experimental parameters, $\beta=1.1972...$, the transition is broadened, and shifted towards larger $\Delta/J$, see Fig.1. Due to the quasi-periodic nature of the potential, these localized states appear approximately every 5 sites (2.6 $\mu$m).

The non-interacting Bose-Einstein condensate is prepared by sympathetically cooling a cloud of interacting $^{39}$K atoms in an optical trap, and then by tuning the s-wave scattering length almost to zero by means of a Feshbach resonance[17,23] (see Methods Summary). The spatial size of the condensate can be controlled by changing the harmonic confinement provided by the trap. For most of the measurements the size along the direction of the lattice is $\sigma\sim$5 $\mu$m. The quasi-periodic potential is realized by using two lasers in standing-wave configuration

**Figure 3 | Observing the nature of the localized states. a, b** Experimental profiles and fitting function $f_\alpha(x)$ (red line) for $\Delta/J=1$ (a) and $\Delta/J=9$ (b). Note the vertical log scale. The blue line represents a Gaussian fit, $\alpha=2$. **c**, Dependence of the fitting parameter $\alpha$ on $\Delta/J$, indicating a transition from a Gaussian to an exponential distribution.

In this regime, the eigenstates of the hamiltonian (1) are *exponentially* localized. We have therefore analyzed the tails of the spatial distributions with an exponential function of the form $f_\alpha(x)=A\exp(-|(x-x_0)/l|^\alpha)+B$, with the exponent $\alpha$ being a fitting parameter. Two examples of this analysis for weak and strong disorder are shown in Fig. 3a. The exponent $\alpha$, shown in Fig. 3b, features a smooth crossover from 2 to 1 for increasing $\Delta/J$, signalling the onset of an exponential localization. We note instead that in the radial direction, where the system is just harmonically trapped, the spatial distribution is always well fitted by a Gaussian function ($\alpha=2$).

Information on the eigenstates of the system can also be extracted from the analysis of the momentum distribution of the stationary atomic states in the presence of the harmonic confinement. The width of the axial momentum distribution P(k) is inversely proportional to the spatial extent of the condensate in the lattice. We measure it by releasing the atoms from the lattice and imaging them after a ballistic expansion.

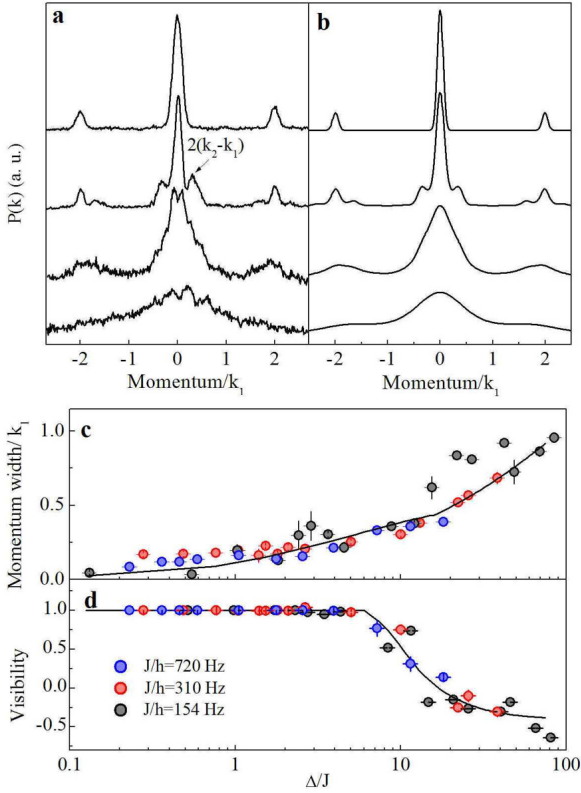

**Figure 4 | Momentum distribution. a,b** Experimental and theoretical momentum distributions P(k) of the BEC released from the quasi-periodic potential for increasing values of $\Delta/J$ (0, 1.1, 7.2, 25, from top to bottom). When $\Delta=0$, we observe the typical interference pattern of a regular lattice. By increasing $\Delta$, we observe peaks at the beating between the two lattices. At the onset of localization $\Delta/J \simeq 6$ the width of the momentum distribution becomes of the order of the Brillouin zone, corresponding to the localization of wavefunction on a single lattice site. The modulation on the top of the density profiles is due to the interference between several localized states. **c**, Rms size of central peak of P(k) for three different values of J, vs $\Delta/J$. The experimental data follow a unique scaling behaviour, in good agreement with the theoretical prediction (continuous line). **d**, Visibility of the interference pattern vs $\Delta/J$. The visibility suddenly drops in correspondence of the localization of the eigenstates on distances smaller that the lattice period at $\Delta/J \approx 6$.

In Fig. 4, we show examples of the experimental momentum distributions with the model predictions, in excellent agreement. Without disorder, we observe the typical grating interference pattern with three peaks at $k=0, \pm 2k_1$ reflecting the periodicity of the primary lattice. The tiny width of the peak at k=0 indicates that the wavefunction is spread over many lattice sites[24]. For weak disorder strength, the eigenstates of (1) are still extended, and additional momentum peaks appear at a distance $\pm 2(k_1 \pm k_2)$ around the main peaks, corresponding to the beating of the two lattices. By further increasing $\Delta/J$, P(k) broadens and its width eventually becomes comparable with that of the Brillouin zone, $k_1$, signalling that the extension of the localized states becomes comparable with the lattice spacing. From the theoretical analysis of the Aubry-André model, we have a clear indication that in this regime the eigenstates are exponentially localized on individual lattice sites. Note that the side peaks in the two bottom profiles of Fig. 4a-b indicate that the localization is non-trivial, i.e. the tails of the eigenstates extend over several lattice sites even for large disorder. In Fig. 4c, we present the *rms* width of the central peak of P(k) as a function of $\Delta/J$ for three different values of J. The three data sets lie on the same line, confirming the scaling behaviour of the system. A visibility of the interference pattern, $V=(P(2k_1)-P(k_1))/(P(2k_1)+P(k_1))$, can be defined to highlight the appearance of a finite population of the momentum states $\pm k_1$, and therefore the onset of exponential localization with an extension comparable with the lattice spacing. In Fig. 4d we show the visibility extracted from the same data above. Experiment and theory are again in good agreement, and feature a sudden drop of the visibility for $\Delta/J \approx 6$.

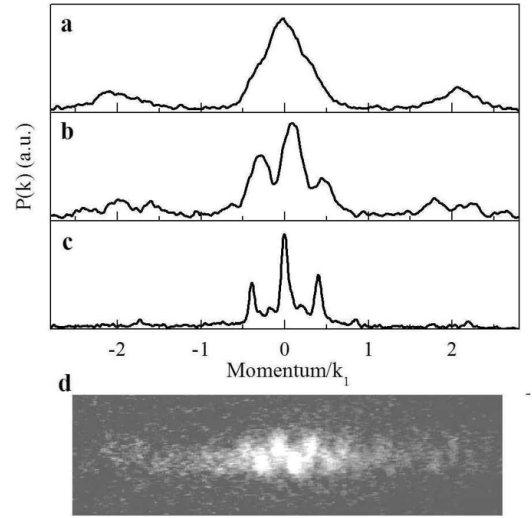

**Figure 5 | Interference of localized states.** Momentum distribution of the condensate prepared in a disordered lattice with $\Delta/J \sim 10$ for different values of the harmonic confinement. **a,** Profile of a single localized state (initial spatial size of the condensate, $\sigma=1.2$ µm), **b,** Interference of two localized states ($\sigma=1.2$ µm), **c,** Three states ($\sigma=2.1$ µm), **d,** Dislocated interference pattern resulting from the interference between two localized states one of which containing a thermally activated vortex.

Further information on the localized states can be extracted from the interference of a small number of them. This regime can be reached in the experiment by simply reducing the spatial extent of the condensate through an increase the harmonic confinement. Typical profiles of P(k) are reported in Fig.5a-c. Depending on the confinement, we can observe one, two or three states, featuring a smooth distribution or a clear multiple-slit interference pattern. The spacing of the fringes yields a spatial separation between the localized states of about 5 sites, as expected. The independent localized states have a quasi-2D geometry, being their axial extent much smaller that the radial one. This feature makes our system an excellent playground to study the physics of quasi-2D systems[25], recently investigated with widely spaced optical lattices[26]. We also observe (Fig. 5d) interference patterns which present a dislocation, possibly produced by thermal activation of a vortex in one of the two localized states[26], in our case for non interacting atoms.

In this work we have observed Anderson localization of coherent non-interacting matter-waves. Future studies might reveal how a weak, controllable interaction affects the observed localization transition. More in general, the high theoretical and experimental control of this system opens a novel scenario for the study of exotic quantum phases arising from the interplay between interaction and disorder [6,8,27].

**Note**. A related work is being carried out in the group of A. Aspect.

**Acknowledgements.** We thank J. Dalibard for stimulating discussions, S. Machluf for contributions and all the colleagues of the Quantum Gases group at LENS. This work has been supported by MIUR, EU (MEIF-CT-2004-009939), INFN, Ente CRF, and IP SCALA.

**Author information.** Correspondence and requests for materials should be addressed to M. I. (e-mail:inguscio@lens.unifi.it).

**Methods**

**Non-interacting Bose-Einstein condensate.** A sample of laser-cooled $^{39}$K atoms is further cooled by thermal contact with $^{87}$Rb atoms in a magnetic trap to about 1μK. The potassium sample is and transferred into an optical trap generated by crossing on the horizontal plane two focused laser beams, and the polarized in the absolute ground state |F=1,m$_F$=1⟩. A further evaporation stage is then performed in presence of a homogenous magnetic field to access a broad Feshbach resonance[17,28]. In this phase the s-wave scattering length is large and positive ($a≈180a_0$, $a_0$=0.529×10$^{-10}$ m) allowing the formation of a stable condensate composed by about $10^5$ atoms at T/Tc<0.1 (Tc=100nK). Once the condensate is produced, we adiabatically bring the magnetic field to 350.0 G, where the residual scattering length is of the order of 0.1 $a_0$ [23]. This corresponds to an atom-atom interaction energy U/J~$10^{-5}$.

**Quasi-periodic optical lattice.** The lattice is created by superimposing two standing waves at incommensurate wavelengths. The primary lattice is generated by a single-mode Yb-YAG laser at 1032nm, whose linewidth and intensity are actively stabilized. The secondary lattice is obtained by a single-mode Ti:Sa laser at 862nm. The bichromatic lattice is mildly focused on the condensate with a beam-waist of about 150 μm. The lattice depths are independently adjusted by means of acousto-optical modulators, and are calibrated by means of Bragg diffraction. The estimated relative uncertainty at 3σ level on the depths is 10%. To prepare the condensate in the quasi-periodic potential, we raise the intensity of the two lattices from zero to the final value in about 100 ms, using s-shaped ramps.

**Analyzing the atomic distributions.** The atomic samples are imaged on a CCD camera through a lens system with a spatial resolution of the order of 5 μm. The images analyzed in Figs.2-3 were recorded in situ, i.e. immediately after release from the trap. The analysis of the profiles shown in Fig. 3, reporting exponential localization, was done by integrating the atomic density distributions along the radial direction. The central 20 % of the signal was then dropped and only the remaining tails were fitted. This procedure reduces possible Gaussian broadening effects due to the transfer function of the imaging system and to the fact that we populate a few localized states. The profiles in Figs. 4-5 were obtained after a long ballistic expansion (25.5 ms) to reduce the contribution of the initial spatial distribution. For the quantitative analysis in Fig. 4c,d, also the width of the spatial distribution was however measured and subtracted from the data. Vertical error bars in Figs.2-5 represent the standard deviation of three to four independent measurements.